# Human-Robot Teams in Entertainment and Other Everyday Scenarios

Pooyan Fazli
Department of Computer Science
University of British Columbia
Vancouver, BC, Canada
pooyanf@cs.ubc.ca

Alan K. Mackworth
Department of Computer Science
University of British Columbia
Vancouver, BC, Canada
mack@cs.ubc.ca

## ABSTRACT
A new and relatively unexplored research direction in robotics systems is the coordination of humans and robots working as a team. In this paper, we focus upon problem domains and tasks in which multiple robots, humans and other agents are *cooperating* through *coordination* to satisfy a set of goals or to maximize utility. We are primarily interested in applications of human robot coordination in entertainment and other activities of daily life. We discuss the teamwork problem and propose an architecture to address this.

## Categories and Subject Descriptors
I.2.9 [**Artificial Intelligence**]: Robotics -- *Commercial robots and application..I*.2.11 [**Artificial Intelligence**]: Distributed Artificial Intelligence -- *Multiagent systems, Coherence and coordination*, *Intelligent agents.*

## General Terms
Design, Reliability, Experimentation, Security, Human Factors, Theory.

## Keywords
Robot-Human-Agent Teams, Robot-Human-Agent Team Coordination and Cooperation, Robot-Human-Agent Planning, Entertainment Industry.

## 1. INTRODUCTION
With advancements in robotic systems, robots are emerging from primarily industrial and medical domains into daily life applications. They are now proposed for use in personal assistance services, elder care, housework, entertainment, education and even as future 'partners' of humans [9].

Of all the emergent technologies of the 21st century, robotics, arguably, is one of the most important. In purely financial terms, robotic entertainment and educational markets reached $184.9 million in 2000 and are anticipated to reach $2.985 billion by 2014 [1]. The service and personal robot market is estimated to go from $600 million in 2000 to $51.7 billion in 2025 [2]. How the market could develop in the future then, is of great interest from several points of view.

Considering the highly dynamic and uncertain environments in which future robots will have to interact compared to that of the familiar and relatively simple industrial robot, the challenge becomes the integration of the advanced physical and cognitive systems required by the next generation of robots. It is not feasible to try to design a 'universal' robot capable of working within a wide range of applications. Indeed, the current entertainment and educational robots such as Aibo [8] and Nao[1] are examples of the existing challenges of cost, long product life cycle and limited functionality; they serve as examples of current limitations which may well apply to future robots as well.

It is generally believed that multi-robot systems hold several advantages over single-robot systems. The most common motivations for developing multi-robot system solutions in real world applications are that a single robot cannot deal with task complexity adequately; the task is spatiotemporally distributed; building several niche, resource-bound robots is easier than building a single powerful robot; multiple robots can support parallelism; and finally, redundancy increases robustness [10]. Moreover, future robotic systems will require robots, humans and other more or less intelligent agents to work in close coordination. In domains such as search and rescue, space exploration, or military operations, teamwork between humans and robots will be required to make a mission safer, faster, reliable and more efficient.

---
[1] http://www.aldebaran-robotics.com/eng/

In this paper, we primarily focus on applications of human-robot teams in everyday life, evaluate the design criteria involved and develop a framework for cooperative systems between humans and their robotic helpers. Considerable progress has been made in multi-robot coordination in various domains, but the major challenge of human-robot teamwork is still an open issue. One-on-one human-system (HCI), human-robot interaction (HRI) and multi-human teams in computer-supported collaborative work (CSCW) are all now established areas of research which mostly rely upon system constraints that are known in advance. Robots and humans in a team must coordinate and adapt their decision making processes and behaviors along with the dynamic requirements and changing expectations of each other. Research done by [12], [5], and [7] are among the notable works in this domain.

## 2. HUMAN-ROBOT COORDINATION IN DIFFERENT DOMAINS
Here we consider human-robot teams in various scenarios.

### 2.1 Movie Industry
In lieu of creating fictional robots through graphics software, or of building robots without cognitive function, it will be possible in the future to have robots, capable both physically and cognitively, of working with humans in movies and live shows. Sophisticated decision making and behavior control mechanisms will be required in order for the robot to deal, as a human must, with such a fluid environment.

### 2.2 Museums, Galleries: Tour Guides or Guards
Robots would be useful in museums or galleries as guides or guards. They would need to interact with visitors and human staff and would need to coordinate the required tasks with humans and with other robots in order to do the job successfully [4].

### 2.3 Mixed Robot-Human Sport Challenge
*2.3.1 Segway Soccer*
Segway soccer, as a contest between two robot-human mixed teams, has been designed to explore issues that would arise in such a situation. Soccer, an adversarial domain, challenges the players to coordinate, integrate and interact [3]. Part of the challenge is the requirement that players maintain a 1.0m safety distance from each other and that both human and robot must interact with the ball prior to scoring a goal.

*2.3.2 Wheelchair Sports*
Wheelchair sports are variations of the able-bodied versions of challenging sports such as soccer and basketball. All of the participants are, for some reason, confined to a wheelchair. A mixed team of humans and robots would be possible to make the game even more accessible, interesting and challenging, using smart autonomous and semi-autonomous wheelchairs.

### 2.4 Health Care Facilities
Hospitals, nursing homes and long term care facilities are suitable environments where robots, acting cooperatively with doctors, nurses and staff, could make an enormous difference in making life easier both for the patients and those who work with them. Human resources are often strained due to task complexity, demographic constraints, or simply funding problems. The domain of health provision has many complex constraints among the tasks and the goals, making tight coordination among the agents essential in order to provide the necessary services and functionalities within the care facility.

### 2.5 Smart Homes
Smart homes will allow the elderly to live independently, safely and with dignity. Indeed, smart homes would make life easier and permit aging-in-place. Health monitoring devices, sensory systems, smart wheelchairs [13] and robots are essential for a smart home. Robots, acting in an intelligent and coordinated fashion would provide activity monitoring, resource management, advice and scheduling. These robots would learn as patterns emerged [11].

## 3. ROBOTS, HUMANS, AND OTHER AGENTS: THE TEAMWORK PROBLEM
The major question in this research is how, in an environment with many heterogeneous unreliable agents, including humans, robots, sensor networks, and other agents, can coordination be achieved to satisfy the mission goals? Designing and constructing heterogeneous teams of robots, agents and humans raises many research challenges, although some of those are present in multi-robot systems [6]. Consider these factors:

- *Task/Role Allocation:* Coordination approaches should consider the different capabilities of robots and humans in task/role allocation. For example, robots are good at computation but humans are good at visual processing.

- *Coalition Formation:* An agent might not have all the resources required to perform a task so a coalition of agents (a mixed human-robot team) might be needed.

- *Communication:* Within the environment, effective communication from robot to robot, robot to human, and human to human is needed.

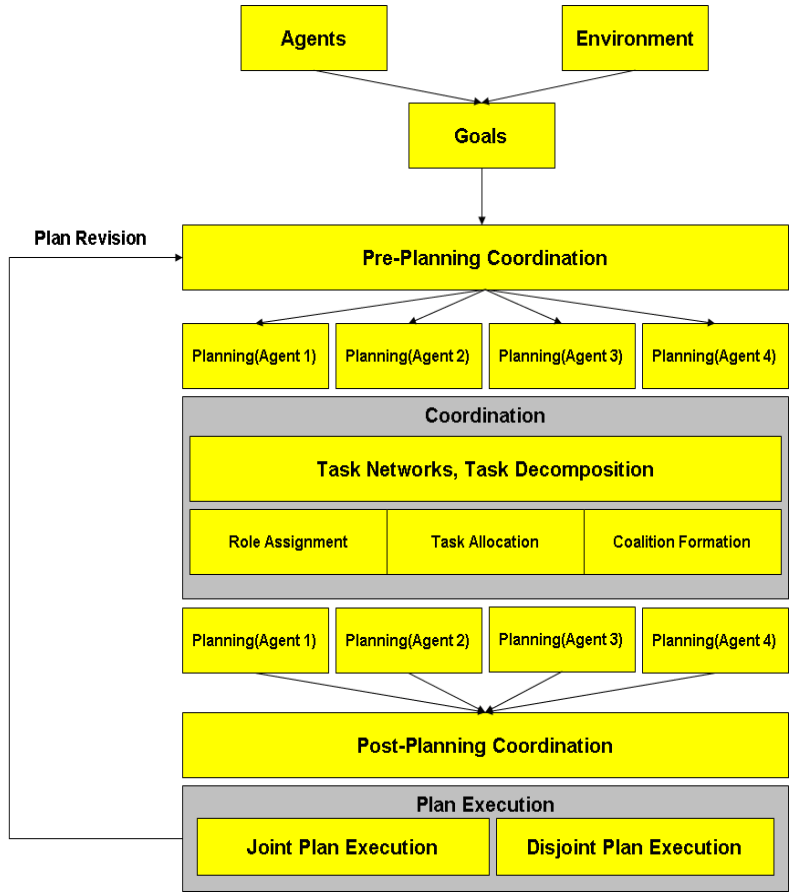

**Figure 1. High Level Overview of the Framework for Exploring Issues of Robot-Human Teams.**

- *Heterogeneity:* The team can consist of robots and humans with different capabilities and performances.

- *Team Integration:* Humans and robots should have mutual understanding and coordinate their behaviors, goals, abilities, and plans, in different situations. The key to this issue is to model humans and robots as an integrated team.

- *Safety and Reliability:* Since robots and humans will be in close proximity, or touching, the safety, reliability and liability issues are crucial.

- *Performance Evaluation:* The system should be able to measure and predict humans and robots performance when interacting with each other.

- *Robustness/Fault Tolerance:* The system should be able to adopt mechanisms to recover from action failure, partial or total robot, human, or agent failure, communication failure, resource failure, and so on.

- *New Input:* New inputs should be handled by the team. For example, what happens if a new robot/human enters the team or a new task enters the list in the middle of the execution when there is no prior model of how these new inputs are generated? New resources and roles can be introduced to the system as well.

- *Uncertainty:* Uncertainty in the environment is an important issue. The team might have a local knowledge of the environment. In unknown or partially known environments as new observations are made, initial solutions may no longer be optimal or efficient. The system should also be capable of handling issues like action uncertainty and noisy sensors.

- *Learning:* The process of learning environment models demonstrates significant differences between robots and humans.

- *Adaptability:* Humans and robots have different abilities in adapting to new situations, such as changes in environmental conditions, objects to be manipulated, and so on.

- *Human Variability:* Humans interact with the environment with variable and unpredictable response times. That variability makes solutions unreliable.

- *Degree of Autonomy:* A distributed decision making process requires various degrees of autonomy. The question is how much authority a robot or a human has for local or global decision making within the team?

## 4. PROPOSED APPROACH

Figure 1 shows the high level overview of the proposed framework in order to explore the issues mentioned in the previous section for highly heterogeneous teams of autonomous robots, humans, and other agents in a multi-goal environment [14].

Our agents (robots, humans, and other components of the domain) work in a multi-goal environment. Each goal (e.g. safety of a wheelchair in a wheelchair soccer game) is a complex task which consists of a set of dependent sub-goals (sub-tasks). Goals are generated by the agents or the environment. To satisfy a goal, all its sub-tasks need to be performed. We want to allocate and schedule real-time tasks, with *precedence* and *time* constraints in a cooperative multi-robot, multi-human setting. An agent can perform a task by itself, or might need to form a coalition of agents. In some scenarios, agents might have conflicting goals too. In such a case, the aim is to maximize the utility for each agent.

Planning is the process of building a sequence of actions that, when executed by the agents in the environment, achieves a given goal. Multi-agent planning coordinates the actions of multiple agents to achieve a goal [14]. Each agent has unique or non-unique control over a subset of resources and can perform different roles based on resource availability. In this framework, task decomposition and task network creation are done in the planning stage. Task networks are dependency networks that show the precedence and time constraints among the tasks to achieve a certain goal. Agents also communicate local plans and coordinate them with other agents; as a result, local plans may be revised which increases robustness and handles part of uncertainties in the environment. In the end, plans can be executed individually or jointly by a team of agents.

## 5. CONCLUSION

Bringing robots from the industrial domain into daily life applications presents many challenges. In this paper, we have presented the concept of coordination and cooperation among teams of robots, humans, and other agents. We introduced the entertainment industry and other daily life scenarios as a potential domain for exploring robot-human teams. We also described some of the significant factors in designing and developing mixed human-robot teams and proposed a teamwork architecture to address these issues in such a complex domain.